\documentclass[aps,prb,twocolumn,showpacs,amsmath,amssymb,floatfix,groupedaddress]{revtex4-1}

\usepackage[utf8]{inputenc}

\usepackage{graphicx}
\usepackage{bm}
\usepackage{color}
\usepackage{epstopdf}
\usepackage{units}
\usepackage{natbib}
\usepackage{ulem}
\begin{document}

\title{Heisenberg scaling with weak measurement: \\A quantum state discrimination point of view}

\author{Andrew N. Jordan$^{1,2}$, Jeff Tollaksen$^{2,3}$, James E. Troupe$^{2,3}$, Justin Dressel$^4$, and Yakir Aharonov$^{2,3,5}$}
\affiliation{$^1$Department of Physics and Astronomy \& Center for Coherence and Quantum Optics, University of Rochester, Rochester, New York 14627, USA}
\affiliation{$^2$Institute for Quantum Studies, Chapman University, 1 University Drive, Orange, CA 92866, USA}
\affiliation{$^3$Schmid College for Science and Technology, Chapman University, 1 University Drive, Orange, CA 92866, USA}
\affiliation{$^4$ Department of Electrical and Computer Engineering, University of California, Riverside, California 92521, USA}
\affiliation{$^5$School of Physics and Astronomy, Tel Aviv University, Tel Aviv, Israel}

\date{\today}


\newcommand{\mean}[1]{\langle #1 \rangle}           
\newcommand{\cmean}[2]{\,_{#1}\langle #2 \rangle}   

\newcommand{\ket}[1]{|#1\rangle}                    
\newcommand{\bra}[1]{\langle #1|}                   
\newcommand{\ipr}[2]{\langle #1 | #2 \rangle}       
\newcommand{\opr}[2]{\ket{#1}\bra{#2}}              
\newcommand{\pprj}[1]{\opr{#1}{#1}}                 

\newcommand{\Tr}[1]{\mbox{Tr}\left[#1\right]}       
\newcommand{\comm}[2]{\left[#1,\,#2\right]}         
\newcommand{\acomm}[2]{\left\{#1,\,#2\right\}}      
\def\R{\mbox{Re}}                                   
\newcommand{\op}[1]{\hat{#1}}                       
\def\prj{\op{\Pi}}                                  

\newcommand{\oper}[1]{\mathcal{#1}}                 
\newcommand{\prop}[1]{\textit{#1}}                  
\def\gbar{\bar{\gamma}}
\def\ebar{\bar{\eta}}
\def\be{\begin{equation}}
\def\ee{\end{equation}}
\def\la{\langle}
\def\ra{\rangle}
\begin{abstract}
We examine the results of the paper ``Precision metrology using weak measurements'', [Zhang, Datta, and Walmsley, arXiv:1310.5302] from a quantum state discrimination point of view.  The Heisenberg scaling of the photon number for the precision of the interaction parameter between coherent light and a spin one-half particle (or pseudo-spin) has a simple interpretation in terms of the interaction rotating the quantum state to an orthogonal one.  In order to achieve this scaling, the information must be extracted from the spin rather than from the coherent state of light, limiting the applications of the method to phenomena such as cross-phase modulation.  We next investigate the effect of dephasing noise, and show a rapid degradation of precision, in agreement with general results in the literature concerning Heisenberg scaling metrology.  We also demonstrate that a von Neumann-type measurement interaction can display a similar effect.
\end{abstract}

\maketitle

In 1988, Aharonov, Albert, and Vaidman\cite{AAV} introduced the concept of a weak value as controlling an anomalously large deflection of an atomic beam passing through a Stern-Gerlach apparatus.  The deflection size is controlled by pre- and post-selected states, as well as the size of the magnetic field gradient.  In the concluding paragraph, they mention that ``another striking aspect of this experiment becomes evident when we consider it as a device for measuring a small gradient of the magnetic field...Our choosing [of the postselection state] yields a tremendous amplification''.  The price one pays for this amplification is the loss of a large fraction of events due to the postselection.  Nevertheless, the relevant information about the parameter in question is concentrated into these small number of events.\cite{me}
This technique has been adapted to optical metrology, and has been successfully implemented in many experiments to precisely estimate various parameters, such as beam deflection, phase or frequency shifts.  For recent reviews of this active area of research, see Refs.~\onlinecite{review1,review2}.

While still obeying the standard quantum limit, weak value amplification experiments have been shown to be capable of extracting nearly all of the theoretically available information about the estimated parameter in a relatively simple way.  Further, it has been shown that in comparison to a standard experimental technique, and given the presence of certain types of noise sources or technical limitations obscuring the measurement process, the weak value-type experiment can have better precision (even when using optimal statistical estimators), even though the detector only collects a small fraction of the light in the experiment.\cite{me}  There have also been a number of recent advances that propose to improve the intrinsic inefficiency of the post-selection.  For example, in the optical context, it is possible to recycle the rejected photons, further improving the sensitivity of the technique.\cite{recycle}  This then gathers all the photons in the experiment through repeated cycles of selection, leading to higher power on the detector with the enhanced signal.  

Quantum-enhanced metrology is based on using quantum resources, such as entanglement, to estimate a parameter of interest better than an analogous classical technique could do with similar resources - typically photon number.  Proposed applications of this field range from precision measurements in optical interferometry to gravity wave detection.\cite{dowling}  Recently, Pang, Dressel, and Brun proposed combining the weak value technique with additional entangled quantum degrees of freedom to further increase the weak value at the same post-selection probability, or to keep the same weak value while boosting the post-selection probability.\cite{entangled}  
This technique leads to Heisenberg scaling of the parameter estimation precision with the number of auxiliary degrees of freedom, using quantum entanglement as a resource.  These advances lead us naturally to consider how other quantum resources manifest in the context of weak measurements, which is the subject of the present article.

An important tool in quantum-enhanced metrology is the Fisher information.  Classically, this quantity indicates how much information about the parameter of interest is encoded in the probability distribution of a random variable that is being measured.  It is an important quantity because it sets the (Cram\'er-Rao) bound for the minimum variance of any unbiased estimator for the parameter of interest.   Any estimator that achieves that bound is said to be efficient.   The quantum mechanical extension of the Fisher information analogously gives the quantum Cram\'er-Rao bound, which indicates the minimum variance achievable using any measurement strategy.  Despite these powerful properties, the formal expressions for the Fisher information do not necessarily provide deeper insight about the physics of the detection method, and can even obscure what are essentially simple physical effects.  Further, just because an estimator has a variance higher than the Cram\'er-Rao bound does not necessarily imply that it is worse that the efficient estimator in practice.  For example, in the case of time-correlated noise, the Fisher information indicates there is a great deal of information about the parameter of interest in the noise-distribution.\cite{me}  However, in order to implement the optimal estimator, extensive post-processing is necessary to properly weight the data points with the auto-correlation matrix elements, adding further resources to the estimation task that are not properly quantified solely by the Fisher information. 

A conundrum involving Fisher information was recently presented by Zhang, Datta, and Walmsley \cite{walmsley}, who considered a coherent state of photons interacting with a spin-1/2 particle in order to estimate a small coupling parameter in the interaction Hamiltonian.  Notably, this example is a variation of the original weak value amplification scenario,\cite{AAV} but using a different parameter regime that more commonly appears in cavity and circuit QED.\cite{cqed1,cqed2}  Even though the coherent state used in their example is typically considered to be a classical quantity that does not provide quantum resources, the authors showed the surprising result that the Fisher information about the coupling parameter seemed to scale at the optimal Heisenberg limit as the average number of photons were increased, rather than at the standard quantum limit that one would typically expect.  This result raises several immediate questions: Is there a simple physical explanation of this apparent Heisenberg scaling, and can this scaling really be used to enhance the estimation of the interaction parameter in an experiment?  

The proposal starts with a separable state of the system $\psi_i$ (a two state system), and a meter state $|\alpha \ra$ (a macroscopic coherent state), given by
\be
| \Psi_0 \ra = [\cos (\theta_i/2) | -\ra + \sin (\theta_i/2) e^{i \phi_i} |+\ra] |\alpha \ra.
\label{in}
\ee
An interaction Hamiltonian generates a unitary operation of the form 
\be
U = \exp (i g \sigma_z {\hat n}), 
\label{interaction}
\ee
which entangles the states.  Here, $\sigma_z$ is a Pauli operator, and ${\hat n}$ is a photon number operator.  This results in the entangled state
\be
| \Psi \ra = \cos (\theta_i/2) | -\ra | \alpha e^{i g} \ra + \sin (\theta_i/2) e^{i \phi_i} |+\ra |\alpha e^{-i g}\ra,
\label{cat}
\ee
which is often called a Schr\"odinger cat state because the total quantum state involves a superposition of macroscopically distinct states of light.

The authors go on to look at projection of the system state onto a final state $\psi_f$, where this state has the same form as $\psi_i$, with the subscript $i$ replaced by $f$ on the parameters.\cite{walmsley}  Specifically, a strong measurement will project the system onto $\psi_f$, or onto the state orthogonal to $\psi_f$ (since the system is two dimensional, there are no other options).  The scaling of the postselected parameter estimation is optimized when pre- and post-selected states are parallel $\psi_i = \psi_f = (|- \ra + |+ \ra)/\sqrt{2}$, so we focus on this case for simplicity of calculation.  The orthogonal state is then clearly $\psi_f^\perp = (|-\ra - |+\ra)/\sqrt{2}$.

In the case of a projection onto $\psi_f$, or $\psi_f^\perp$, the resulting meter states of the light are given by
\be
| \phi_{\pm}\ra = (1/2) ( | \alpha e^{i g}\ra  \pm |\alpha e^{-i g}\ra),
\ee
where $+$ refers to projection onto $\psi_f$, and $-$ onto $\psi_f^\perp$.
These meter states must be properly renormalized, which gives the probability $p_\pm$ of projecting on the parallel or perpendicular system states,
\be
p_{\pm} = 1/2 \pm (1/4) (\exp( |\alpha|^2 (e^{2 i g} - 1))+c.c.).
\label{prob}
\ee
Note that if $g \rightarrow 0$, the probability to project back onto the initial system state limits to 1. Ref.~\onlinecite{walmsley} points out that there are three possible sources of information in the measurement: the probability of the postselection projection $p_{+}$, and the information in the two meter states, $| \phi_{\pm}\ra$ (in principle, the correlations between these outcomes also have information in them).   The Fisher information contained in these channels is then calculated, and curiously, while the meter states have Fisher information that scales with $N = |\alpha|^2$ (yielding the standard quantum limit), the probability of the \textit{post-selection} has a Fisher information that scales as $N^2 = |\alpha|^4$, giving \textit{Heisenberg} scaling in the photon number for the precision of estimating $g$.

The main purpose of this paper is to give physical insight into why Heisenberg scaling for the parameter $g$ can be obtained at all, and further, why it comes mainly from the probability of projecting on the system state, as opposed to mining the meter states for information, as is usually done in weak value amplification experiments.\cite{review1}  Zhang, Datta, and Walmsley write that ``How this conditioning step using a classical measurement apparatus achieves a precision beyond the standard quantum limit is therefore an interesting open question.''
We answer this question here, and give a simple physical argument showing how this scaling is possible.

{\it Approach}.---We approach the question by mapping the problem of obtaining a precise estimate for the parameter $g$ onto a different problem:  Under what conditions can one distinguish the entangled state $| \Psi \ra$ from the separable state $|\Psi_0\ra$?  It is well-known in quantum physics that two states can only be reliably distinguished if they are orthogonal to one another.\cite{book}   Therefore, $g$ must be large enough to move the initial separable state to an orthogonal state.  This sets the scale of the minimum value for $g$ that can be reliably distinguished when using the two-state system as a probe.  Unless the states are distinguishable, no processing techniques will help in the metrological task.

{\it Spatial shift of independent meter states.}---
We begin with first illustrating this principle on a system exhibiting standard quantum limit scaling.  Consider a photon prepared in a Gaussian wavefunction, of zero mean and width $\sigma$.  The wavepacket interacts with a two-state system, and is shifted in position by a distance $\pm d/2$, depending on which state the two-state system is in.  This is described by a standard von Neumann type interaction as usually found in measurement models,
\be
U_{v} = \exp(i d {\hat p} \sigma_z/\hbar).
\label{vn}
\ee
Here $ {\hat p}$ is the momentum operator of the meter.
If the system begins in the separable state $|\Phi_0\ra=(1/\sqrt{2})(|+\ra + |-\ra) \psi_0(x)$, the interaction results in the state $\la x|\Phi\ra = (\psi_+(x) |+\ra + \psi_-(x) |-\ra)/\sqrt{2}$.  Here, $\psi_\pm(x) = (2\pi \sigma^2)^{-1/4} \exp[-(x \pm d/2)^2/4\sigma^2]$, and $\psi_0 = \lim_{d\rightarrow 0}\psi_\pm(x)$.  Taking the overlap ${\cal O}$ between state $|\Phi\ra$ and the original separable state $|\Phi_0\ra$, we find ${\cal O} = \la \Phi_0 | \Phi \ra = \exp[-d^2/32 \sigma^2]$.  For $N$ independent photons, this overlap is raised to the $N^{th}$ power because the state is simply a product of one photon states.   The smallest value of $d$ that can be measured is when this overlap is nearly zero, which corresponds to $d_{min} \sim 4 \sigma/\sqrt{N}$.

This result for a minimum resolvable position can be compared with the quantum Fisher information\cite{Caves} in the state $\Psi(g)$ about the parameter $g$.  The quantum Fisher information for pure states is defined as 
\be
F(g) = 4 \frac{d \la \Psi|}{dg}  \frac{ d|\Psi\ra}{dg}  - 4 \left\vert \frac{d \la \Psi|}{dg} |\Psi\ra\right\vert^2.
\label{FI}
\ee
Applied to state $|\Phi\ra$ above with $g$ identified with $d$, and generalizing to $N$ independent photons, we find the result,
\be
F = \frac{N}{4 \sigma^2}.
\ee
This Fisher information sets the minimum resolution on the detectable value of $d$, the quantum Cram\'er-Rao bound, $d_{min} = F^{-1/2} = 2 \sigma/\sqrt{N}$.   This result is the same as the classical Cram\'er-Rao bound from the probability distributions $P_{\pm}(d) = |\psi_{\pm}|^2$, and coincides with the standard quantum limit scaling with $N$.  We therefore see that both the state overlap criterion and the Fisher information approach give similar results.

{\it Cavity QED interaction.}--- We now return to the situation described in the introduction, and reconsider that situation in light of the state overlap distinguishability criterion.  Computing the overlap ${\cal O}$ between $| \Psi_0\ra$ (\ref{in}) and $ | \Psi \ra$ (\ref{cat}), we find
\be
{\cal O} = \la \Psi_0 | \Psi \ra = (1/2) (\exp( |\alpha|^2 (e^{ i g} - 1)) + c.c.).
\label{overlap}
\ee
In order to investigate the conditions under which this expression can decay to zero, we replace $|\alpha|^2 = N$, the average number of photons in the coherent state, and consider a small $g \sim 1/N$: Heisenberg scaling on the precision of $g$.  

We expand the terms in the exponential in the above overlap in powers of $g$ to obtain to leading order,
\be
{\cal O} = e^{- N g^2/2} \cos g N.
\label{o}
\ee
Thus, by passing $g \sim 1/N$ through the zero of the cosine, the state can be made orthogonal; the exponential suppression is negligible for such a small $g$.  We conclude that one can indeed distinguish a $g$ of order $1/N$.  Plots of the exact overlap and the projection probability are given in Fig. 1.

\begin{figure}
\includegraphics[width=8cm]{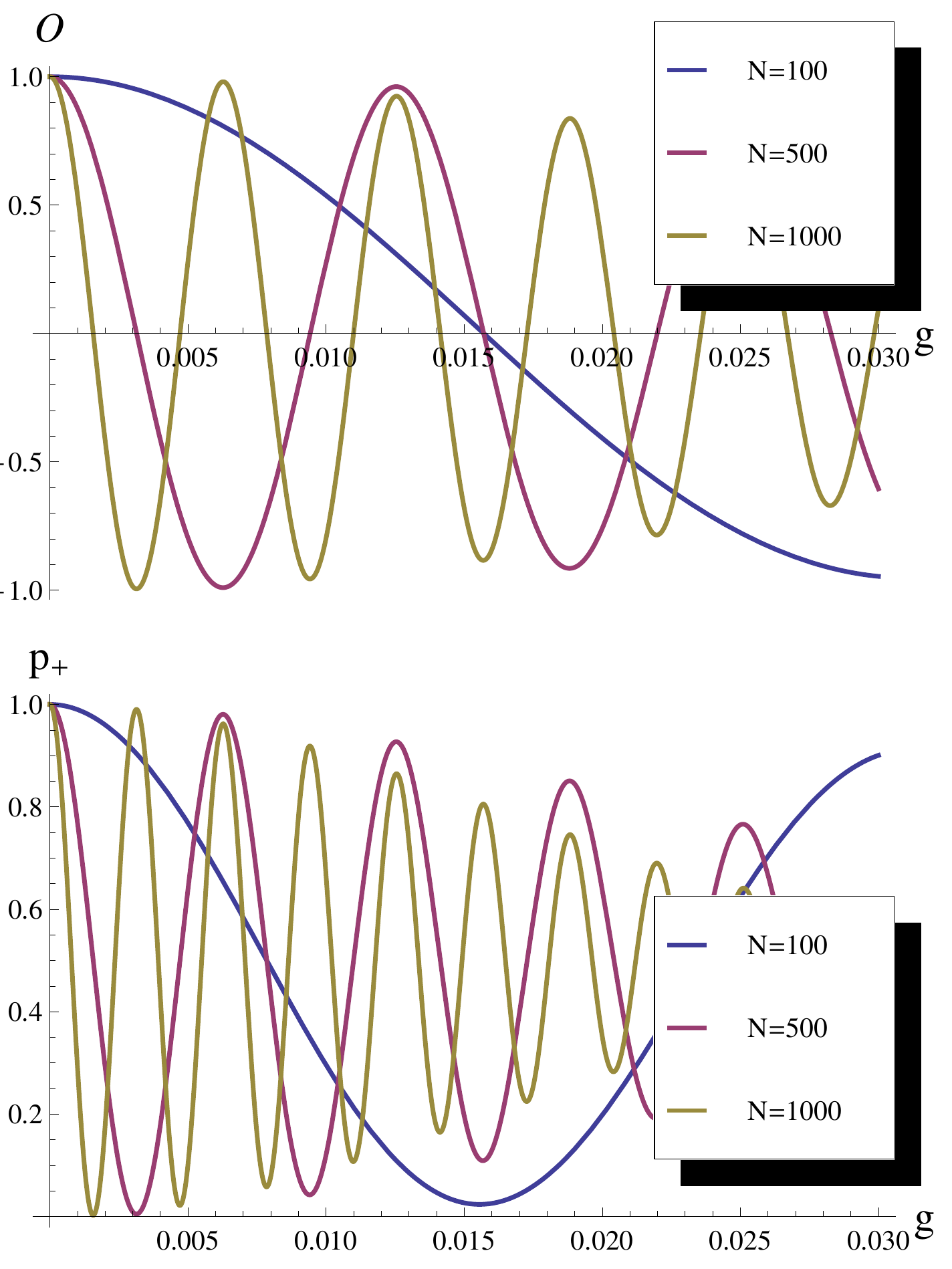}
\caption{The overlap (top) and postselection probability on the $|+\ra$ state (bottom) are plotted versus the interaction parameter $g$ for different values of $N$, the mean photon number in the coherent state.  As $N$ increases, both quantities oscillate more quickly with $g$, permitting better discrimination of the value of $g\sim 1/N$.}  
\label{fig-xyz}
\end{figure}

One can understand why the overlap can be made to vanish because for large $N$, and $g$ of order $1/N$, the perturbation to the coherent state is effectively an $N$-independent phase shift of the coherent state, which combined with the system states in $|\Psi\ra$, simply produces a limiting state of,
\be
| \Psi'\ra \approx (1/\sqrt{2}) (|+\ra e^{i g N} + |-\ra e^{- i g N}) |\alpha\ra + |\psi'\ra|\alpha_\perp \ra,
\label{after}
\ee
where $|\alpha_\perp\ra$ is a state orthogonal to $|\alpha\ra$ and $|\psi'\ra$ is another system state.  The fact that $|\alpha_\perp\ra$ is orthogonal to the initial state makes it (and  $|\psi'\ra$) irrelevant to the orthogonality criterion.
Thus, the nonlinear interaction has the effect of just rotating the spin, while leaving the coherent state untouched, at least in the large $N$ limit, appropriately scaled.  Since the phase rotation of the qubit that is induced by each photon accumulates \textit{coherently}, the scaling with photon number shows a quantum advantage compared to the incoherent photon accumulations that lead to the standard quantum limit.  

Now we can see why the information about the parameter $g$ is mainly found in the probability of the selection, $p_{\pm}$ in Ref. \onlinecite{walmsley}.  We notice that up to factors of 2 and shifts by constant factors, the selection probability (\ref{prob}) has the same form as the overlap between entangled and initial state (\ref{overlap}).  As we see above, the main effect is the coherent phase rotation of the system state, and consequently, this is the effect that will give a large change of the selection probability that can be used to deduce the value of $g$.   With this insight, it makes perfect sense that the post-measurement meter states have relatively little information that can be extracted, and the Heisenberg scaling appears only in the post-selection probability.

\begin{figure}[tbh]
\includegraphics[width=4cm]{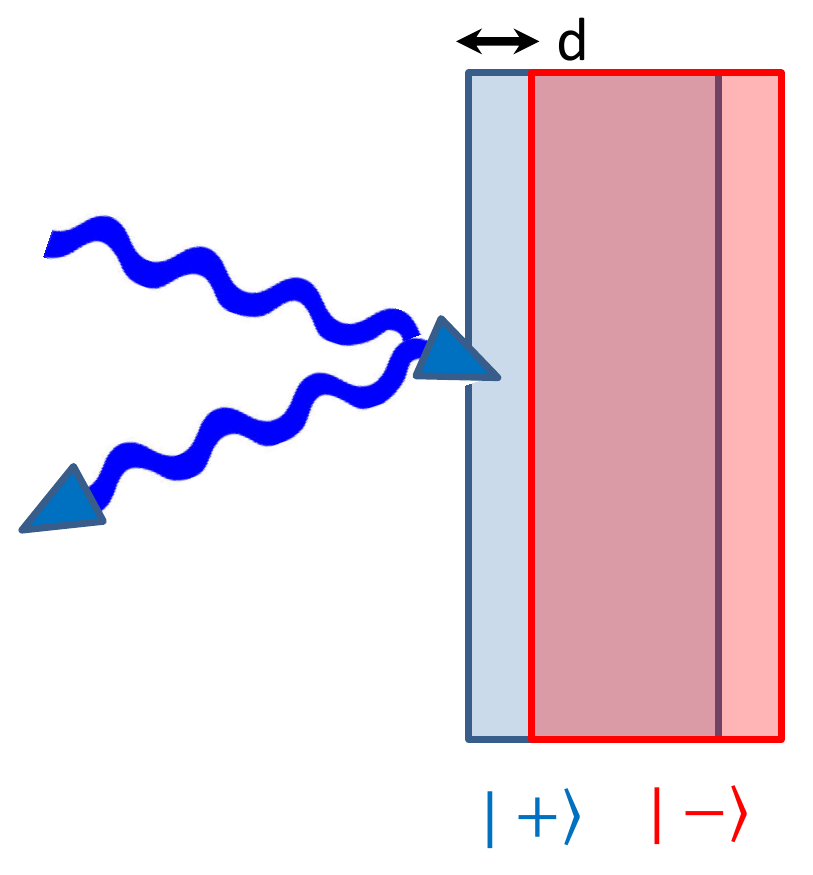}
\caption{Sketch of an experiment to determine the position of a quantum mirror using a coherent light beam.}
\label{fig-yakir}
\end{figure}

From the above analysis, the origins of the Heisenberg scaling of the estimate of the $g$ parameter are made clear; however, there are still important question about the utility of this technique.  For example, can this method be used to measure information about the spin?  We can put this question differently by considering the thought experiment sketched in Fig.~\ref{fig-yakir}.  There, a coherent state is reflected off a quantum mirror (similar to a quantum beam splitter \cite{ionicioiu}).  The mirror location is located in one of two positions, separated by a distance $d$, and described by states $|+\ra, |-\ra$.  The interaction will give a relative phase shift to the beam of $g = d/\lambda$, where $\lambda$ is the wavelength of the light.  In this Heisenberg scaling limit, the fact that the coherent state is unchanged after the interaction (\ref{after}), indicates that it contains no information about the spin (or in this case, the location of the quantum mirror).  Only the mirror states are affected because the mirror itself collects all the phases, and the quantum mirror must itself be probed by measuring its position to find the distance $d$.  This indicates that one cannot use the coherent state light alone to measure the mirror's position at this level of precision. 

{\it Implementations.---} There can be, however, other uses of this technique.  In cross-phase modulation, a single photon, prepared in a superposition of two polarizations (for example) can interact nonlinearly with a coherent beam with a large average photon number.  Depending on the polarization state of the single photon, the phase of the coherent beam is changed by different amounts.  This nonlinearity is very difficult to create optically, and single-photon ``cross-Kerr'' nonlinearities have not been seen yet in the lab.  However, Feizpour, Xing, and Steinberg have shown that in some cases, a single photon can be made to act like many through a weak value amplification process.\cite{FXS}  In this case of cross-Kerr interaction, the difference of rotation angle for the two polarizations is identified with the $g$ parameter, and the method of measuring the frequency of projection of the polarization back on the original state can estimate that parameter. The standard error on $g$ will scale as $1/N$ (the Heisenberg limit in photon number), times $1/\sqrt{\nu}$, where $\nu$ is the number of projections on the two-state system (the standard quantum limit in measurement realizations).  Thus, in order for this technique to be useful, we need the number of photons $N$ per projection to be large, and need to be able to repeatedly measure the single photon polarization.  Measuring the changes in the coherent state of the light is irrelevant for the Heisenberg scaling on the precision of $g$.  Unfortunately, linear optics is unable to realize an interaction of the form Eq.~(\ref{interaction}), and will instead create products of coherent states.  Thus, any interferometric set-ups will be unable to create this state.  In order to create the state, nonlinear methods are needed.  We note, however, that since we can ignore the state of the light entirely (since we know the average photon number $N$), the relevant state is not a Schr\"odinger-cat state, since the coherent state can be traced out entirely, leaving just the phase-shifted qubit state to work with.

A more realistic implementation of the interaction (\ref{interaction}) is in the field of cavity or circuit QED.  There, a superconducting quantum bit (artificial atom) interacts coherently with a microwave field inside a cavity, producing exactly this interaction, called a light shift, or ac stark shift.  See Refs.~\onlinecite{cqed1,cqed2} for recent experiments using this interaction.  This name ``light shift'' is related to a Hamiltonian of the form
\be
H = \hbar \omega_r a^\dagger a + \hbar \omega_a \sigma_z/2 + \hbar \chi \sigma_z a^\dagger a,
\ee
where we have added in Hamiltonian terms for the qubit and the light field.  This Hamiltonian is valid in the dispersive limit, where the detuning between the cavity frequency, $\omega_r$, and the qubit frequency $\omega_a$ is much larger than the microscopic coupling constant of a Jaynes-Cummings interaction.   The interaction can be interpreted as a qubit-state-dependent shift of the cavity frequency, $\chi \sigma_z$, or as an intracavity photon-number-dependent shift of the qubit frequency, $2 \chi a^\dagger a$.  It is the later interpretation that is directly related to the effect we have been discussing. 

As usual, in order to get the unitary development (\ref{interaction}), the two systems are coupled impulsively in time.
The shift of the qubit frequency causes a precession in $x-y$ plane of the Bloch sphere that can be read out via projective qubit measurements, following unitary rotations.  In fact, this qubit frequency measurement is used routinely to determine the average photon number in the cavity, since the parameter $g = \hbar \chi$ can be independently determined spectroscopically as the shift in resonance frequency of the coupled cavity.  In our case, we are interested in the reverse procedure:  A large, known photon number $N$ in the cavity creates a finite phase shift on the qubit for an unknown small coupling parameter $g$ of the order $1/N$.  Hence, measuring the qubit would allow one to independently determine $g$ without using the spectroscopy of the cavity.

The difficulty in implementing the scheme outlined above is the fact that, usually, the way to measure the qubit state is with the cavity field itself, which we already showed becomes uncorrelated with the qubit precisely when the qubit is most sensitive to $g$.  Consequently, the implementation requires that we first have an a weak unknown interaction of the qubit with one cavity, followed by a strong interaction with another cavity to do the projective measurement.  This could be accomplished perhaps with two strip-line resonators, both of which are coupled to a single transmon qubit.

\begin{figure}[tbh]
\includegraphics[width=8cm]{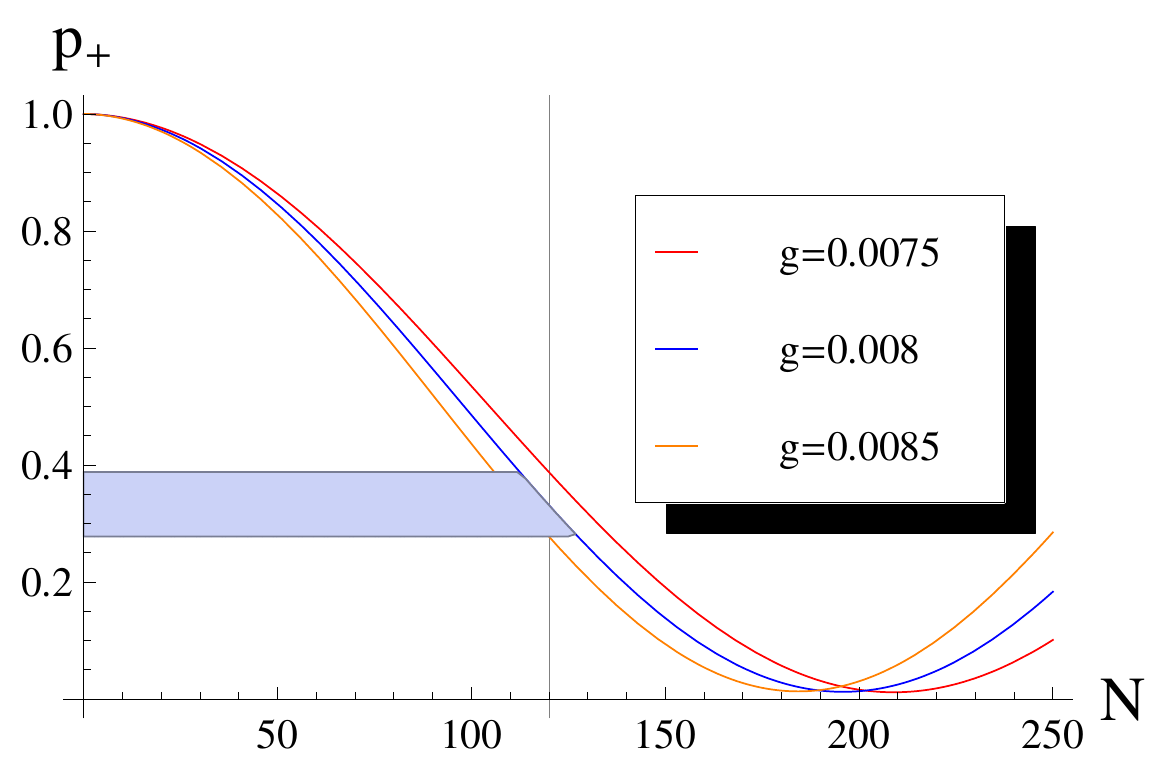}
\caption{Converting uncertainty about $p_+$ into uncertainty about $g$, given a known $N=120$.}
\label{fig-p+}
\end{figure}
Let us now make an analysis of the number of projections needed in a practical series of experiments.  Suppose there is some small, but unknown $g$.   We know the precise average number of photons $N$ in the coherent state.  We furthermore have calculated the probability distribution of the successful post-selection (\ref{prob}) is given approximately by $p_+  \approx  \cos^2 g N$ for small values of $g$.  Depending on the value of $N$, ranging from $0$ to $\pi/2g \gg 1$ the probability will be between 0 and 1 (we assume $g$ is sufficiently small the multiple solutions of the cosine inverse are not relevant).  Suppose we make $\nu$ experiments with a small interaction $g$, followed by a projective qubit measurement, keeping $N$ the same in every experiment.  This will result in $\nu$ binary results of being projected into the initial state, or the orthogonal state, from which we can estimate the probability $p_+$ as simply the number of times the initial state is found, divided by the total.  The experiments are uncorrelated, so the uncertainty on the value of $p_+$ is simply $\sigma_{p_+} = 1/\sqrt{\nu}$.  This uncertainty on the value of $p_+$ then sets the uncertainty $\sigma_g$ on the estimated value of $g$, 
\be
\sigma_g = \left |\frac{\partial p_+}{\partial g} \right |^{-1} \sigma_{p_+} = \frac{1}{a N\sqrt{\nu}},
\ee 
where $a = \sin(2 \cos^{-1} \sqrt{p_+})$ is typically of order 1.  This is the scaling discussed in Ref.~\onlinecite{walmsley}.  The procedure discussed above is graphically represented in Fig.~\ref{fig-p+}.

One might wonder if the two-state system could be dispensed with entirely since the effect comes from the coherent state $| \alpha e^{\pm ig}\ra$.  
We can see this is not so from two perspectives.  In the first approach, the orthogonality criterion gives,
\be
\la \alpha | \alpha e^{\pm i g}\ra = \exp[|\alpha|^2 (e^{i g} -1)].
\ee
Making an expansion for small $g$, $e^{i g} -1 \approx i g -g^2/2 + \ldots$, we see the first term is just an overall phase shift on the coherent state which is not distinguishable on measurements on the state of light.  We must go to the second order terms to obtain orthogonality, which corresponds to $g_{min} \sim 1/\sqrt{N}$, recovering the standard quantum limit for the photon number scaling in the light states.

In the second approach, we can find the quantum Fisher information (\ref{FI}) about $g$ in the light state $| \alpha e^{\pm i g}\ra$.  We find the result $F= 4N$, so the quantum Cram\'er-Rao bound is $g_{min} = F^{-1/2} = 1/\sqrt{4 N}$, which is consistent with the approach above, as well as the standard quantum limit scaling expected from a coherent state.

{\it Effects of dephasing.---} One outstanding challenge to quantum metrology is the fact that in the presence of small amounts of dephasing noise, the Heisenberg scaling rapidly changes to standard quantum limit scaling.   Here, there is some advantage in the sense that photon loss will not have much effect on the coherent state (other than simply lessening the overall magnitude, $N = |\alpha|^2$).  However, as we already showed, the Heisenberg scaling occurs by measuring the two-state system, so this is not really helpful.  The important effect is how the scaling depends on fluctuations on the phase shift that is being measured, $g$.  Phase fluctuations (or other dephasing mechanisms) will then be the most serious detriment to this method, rendering it useless for estimating $g$ better than the standard quantum limit.  Similar difficulties can be seen with N00N states.\cite{dowling2}

We can see this effect by considering the possibility of also acquiring a small, random, phase shift $\phi$ that will be averaged over, so we have $g \rightarrow g+ \phi$.  We take for simplicity that $\phi$ is a Gaussian random variable with zero mean and variance $\la \phi^2\ra$.  The averages involved for the general cat state (\ref{cat}) are needlessly complicated, so we focus instead on the approximate state after scaling, Eq.~(\ref{after}).  In order to carry out the averaging, we need a density matrix representation,
\begin{eqnarray}
\rho &\approx& \frac{1}{2}( |+\ra\la + | + |-\ra \la - |    \nonumber \\
&+& e^{2 i g N + 2 i N \phi} |+\ra \la -|  +   e^{-2 i g N - 2 i N \phi}|-\ra \la + |).
\end{eqnarray}
Here we have neglected entirely the coherent state since it is effectively separable in the Heisenberg scaling regime.  We also neglected the orthogonal contributions since they do not contribute to the overlap.
Taking averages over $\phi$, we are left with a mixed state,
\begin{eqnarray}
\rho' &\approx& \frac{1}{2}( |+\ra\la + | + |-\ra \la - |    \nonumber \\
&+& e^{2 i g N - 2 N^2 \la\phi^2\ra} |+\ra \la -|  +   e^{-2 i g N - 2 N^2 \la\phi^2\ra}|-\ra \la + |).
\label{rho'}
\end{eqnarray}
We can now compute the overlap of this mixed density matrix with the original state $|\psi_i\ra = (|+\ra + |-\ra)/\sqrt{2}$, giving the square overlap, ${\cal O}^2 = \la \psi_i | \rho' | \psi_i \ra$.  We find,
\be
{\cal O}^2  = (1/2) \left(1 + \cos (2 g N) e^{-2 N^2 \la \phi^2\ra}\right).
\ee
In the limit of no noise, we recover (\ref{o}) to leading order, after using the half-angle formula.

In contrast to the noiseless case, we now see that even a small amount of phase noise can destroy the Heisenberg scaling due to the exponential damping that scales as $N^2$ - eliminating the oscillatory behavior that permitted us to estimate $g$.
 The same effect that leads to sensitive estimation of $g$---namely the coherence of the phase rotations of the qubit---is also the source of fragility of the technique.  We can explore this effect from another point of view by calculating the quantum Fisher information $F$ in the state (\ref{rho'}).    This sets the minimum uncertainty on the parameter $g_{min} \ge 1/\sqrt{F}$, the quantum Cram\'er-Rao bound.  Defining $L$ as the logarithmic derivative of the density matrix $\rho'$ with respect to $g$, $F$ for mixed states is defined as 
\be
F = {\rm Tr}(\rho' L^2) = 4 N^2 e^{-4 N^2 \la \phi^2\ra}.
\ee
Thus, we see that the $N^2$ contribution to the Fisher information is quickly degraded in the presence of finite dephasing noise, even for moderate values of $N$.  The Fisher information perspective gives a complementary point of view to the state distinguishability criterion discussed above.

{\it Von Neumann measurement revisited.---} Before concluding, we point out that the effects described with the cavity QED-type interaction (\ref{interaction}) can also be seen more easily with the von-Neumann interaction (\ref{vn}).  Rather than use a meter wavefunction that is Gaussian as is usually considered to extract information about the qubit state, we consider a meter wavefunction of the form of a plane-wave of wavelength $\lambda$, that is $\psi(x) = \exp(2\pi i x/\lambda)$.  Of course, this plane wave should as usual be normalized, either by putting on a slowly decaying envelop function, or with box normalization.   Nevertheless, for this discussion, it is simpler to keep it unnormalized, and this will not change the main conclusions.  We can equivalently write this state in momentum space as $\phi(p) = \delta(p-p_0)$, where the momentum $p_0 = 2 \pi \hbar/\lambda$.  

Starting with the separable state $|\Phi'_0\ra=(1/\sqrt{2})(|+\ra + |-\ra) \phi(p)$, the interaction (\ref{vn}) develops the state to 
\begin{eqnarray}
\la p| \Phi'\ra &=& e^{i d p \sigma_z /\hbar}  \delta(p-p_0) \frac{|+\ra + |-\ra}{\sqrt{2}}, \\
&=&  \delta(p-p_0) \frac{e^{i d p_0 /\hbar}  |+\ra + e^{-i d p_0 /\hbar} |-\ra}{\sqrt{2}}.
\end{eqnarray}
If a momentum measurement is now made on the meter state, post-interaction, we will of course find it in exactly the state we put it in, with momentum $p_0$, giving precisely no information about the state of the qubit.  However, as before, the interaction has rotated the qubit by a phase shift $d p_0/\hbar$.  This procedure may now be repeated for $N$ independent photons, all prepared in state $|p_0\ra$.  The interaction will develop the state in precisely the same way, and the phase shift will simply add, giving a new phase shift of $N$ times the single particle phase shift.  Projecting the qubit back on its original state will find that state with probability
\be
P_{+,N} = \cos^2(d p_0 N/\hbar) =  \cos^2(2\pi d N/\lambda).
\ee
In contrast to the Gaussian meter case, we can now distinguish the distance $d_{min} \sim \lambda/N$, provided we add in the projective measurement possibility on the qubit following the interaction.  It is instructive to see that in this case, the weak measurement is actually no measurement at all, but only a weak interaction affecting the qubit only.  The same is true in the case of the interaction (\ref{interaction}), provided $g$ is sufficiently small.

Given this insight, could this scheme could be implemented in the optical experiments demonstrating weak value amplification by simply monitoring the post-selection probability?  A von Neumann type interaction has been shown using both polarization \cite{pol} or which-path \cite{wp} degrees of freedom.  The answer is that those experiments use a single photon as both the meter (transverse deflection) and system (polarization or which-path), so the number of meter photons per system projection is $N=1$.  

{\it Conclusions.---} 
We have considered the weak measurement metrology model proposed by L. Zhang, A. Datta, I. A. Walmsley,\cite{walmsley} and have given a simple interpretation of the Heisenberg scaling of the Fisher information with photon number shown there:  The coherent state interacting nonlinearly with a spin 1/2 particle imparts a coherent phase shift to the spin that can rotate the spin to an orthogonal state for $g \sim 1/N$.  Unfortunately, the coherent state carries only information about $g$ that scales with the standard quantum limit, and the spin (or pseudo-spin) must be measured directly in order to obtain Heisenberg scaling precision.  
We have further investigated dephasing effects on the scheme, and shown a rapid degradation to the measurement precision, emphasizing that it is the fragile quantum coherence of the spin that leads to the enhanced scaling.  This behavior has been argued to be generic to all Heisenberg scaling schemes (see Refs. \onlinecite{hs}), so it is not surprising that we also find this behavior here as well.  We also showed that the von Neumann measurement interaction also has the phase accumulation effect, provided we prepared the meter states in momentum eigenstates.

{\it Acknowledgments.}—We thank L. Zhang, A. Datta, I. A. Walmsley, and A. N. Korotkov for discussions. Support from the US Army Research office Grant No. W911NF-09-0-01417 and Chapman University is gratefully acknowledged.

\end{document}